\documentclass[aps,prb,showpacs,twocolumn,amsmath,amssymb,floatfix]{revtex4}
\usepackage{graphicx}
\usepackage{Jonasmacros}
\usepackage{bm}
\begin{document}
\preprint{}
\title{Subnanosecond switching of local spin-exchange coupled to ferromagnets}
\author{Jonas Fransson}
\email{Jonas.Fransson@fysik.uu.se}
\affiliation{Theoretical Division and Center for Nonlinear Studies, Los Alamos National Laboratory,
Los Alamos, New Mexico 87545 }
\affiliation{Department of Physics and Materials Science, Uppsala University, SE-751 21\ \ Uppsala, Sweden}

\begin{abstract}
The dynamics of a single spin embedded in a the tunnel junction between ferromagnetic contacts is strongly affected by the exchange coupling to the tunneling electrons. Using time-dependent equation of motion for the spin under influence of the spin-polarized tunneling current, it is shown that the magnetic field induced by bias voltage pulses allows for sub-nanosecond switching of the local spin and the possibility of spin reversal is  illustrated. Furthermore, it is shown that the time-evolution of the Larmor frequency sharply peaks around the spin-flip event, and it is argued that this feature can be used as an indicator for the spin-flip.
\end{abstract}
\date{\today}
\maketitle

Detection and manipulation of single spins is an important field of science since it pushes the limits of quantum measurements. Single spins would be an object for qubits and would, thus, be crucial for quantum information technology. Potentially, spintronics will replace conventional electronics devices with spin analogues where manipulation, control, and read-out of spins enable functionality with no or little charge dynamics.\cite{awschalom2002} Of major importance is to understand how spins can be manipulated through electric fields, and questions of the time-scales involved.

Current-induced magnetic switching have been addressed for planar \cite{slonszewski1996,berger1996,krivorotiov2005,edwards2005,nozaki2006} and magnetomechanical \cite{kovalev2005} systems, as well as questions concerning spin-transfer torque\cite{liu2003,bauer2003,ji2003,ozyilmaz2003} and decoherence.\cite{katsura2006} Recently, the dynamics of a local spin coupled to superconducting leads was considered and found non-trivial in the sense that a finite charge current introduces a nutation of the spin motion.\cite{zhu2004,nussinov2005} Analogous nutations were found for a local spin embedded in the tunnel junction between ferromagnetic leads biased with a harmonic voltage.\cite{fransson2007} Concerning single spin manipulations, such kind of nutation provides significant implications for electrically controlling the dynamics of local spins. Spin-flip transitions of tunneling electrons generated by spin-orbit coupling is also a source for direct manipulations of the local spin.\cite{zhu2006} Spin nutations and spin echo were observed under influence of external microwave magnetic field using force detection techniques for studies of electron-spin resonance.\cite{wago1998}

The purpose of this paper is to consider some implications of the nutations of the local spin that are generated by a time-dependent bias voltage across the junction.\cite{fransson2007} It is argued that that the nutations of the local spin can be electrically controlled and its amplitude can be made sufficiently large for the spin to undergo spin-flip transitions, by application of short bias voltage pulses across the junction, e.g. pulses of time span less than 1 ns. In an arrangement of ferromagnetic leads with unequal magnetization and/or in non-collinear alignment, the local spin tends to line up along the magnetic moment of the source lead, and by reversing the bias across the junction stimulates reversal of the local spin. In this respect, the spin-polarized current can be considered to be generating an anisotropy field that stabilize the orientation of the local spin, as long as the current flows across the junction. It is further argued that the time-evolution of the spin Larmor frequency sharply peaks around the event of the spin-flip transition. This feature can thereby be used as a signal for detection of the spin-flip event.

\begin{figure}[b]
\begin{center}
\includegraphics[width=6.5cm]{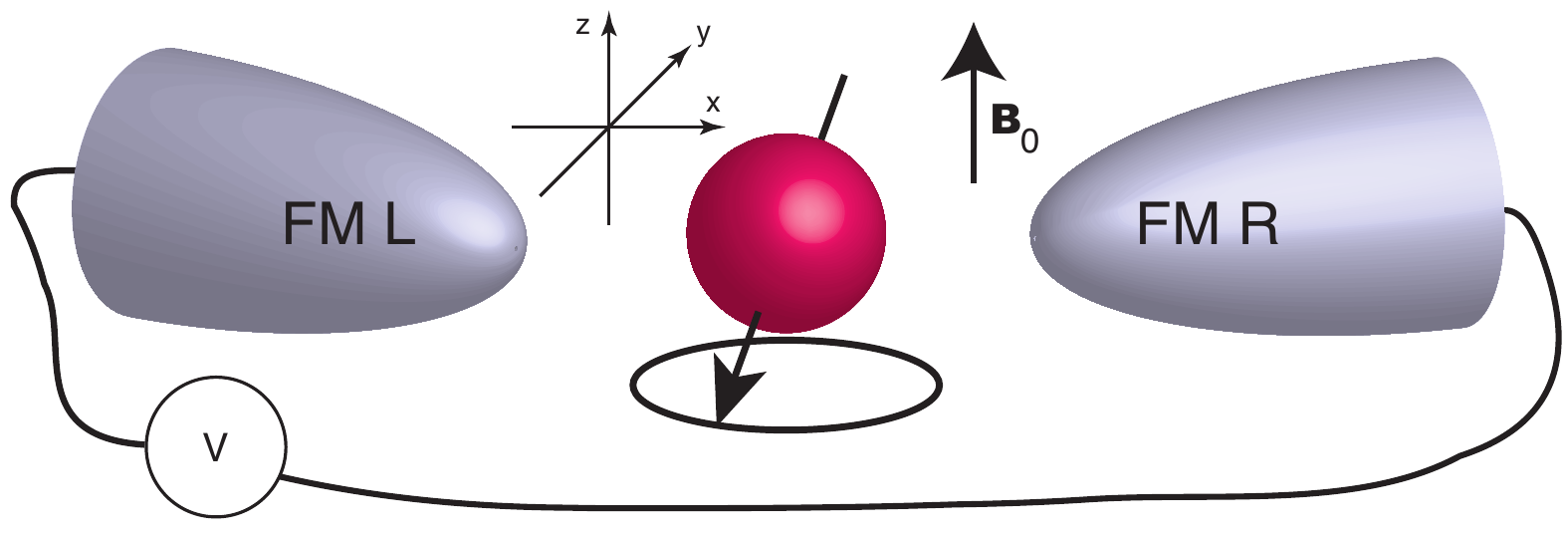}
\end{center}
\caption{(Color online) Local spin embedded in the tunnel junction between two ferromagnetic leads.}
\label{fig-system}
\end{figure}
Following the setup given in Ref. \onlinecite{fransson2007}, a local spin $\bfS$ embedded in the tunnel junction between two ferromagnetic leads, see Fig. \ref{fig-system}, is considered. The leads are directly coupled through tunneling and indirectly through exchange interaction with the local spin. The Hamiltonian for the system can then be written as
\begin{equation}
\Hamil=\Hamil_L+\Hamil_R+\Hamil_S+\Hamil_T.
\label{eq-system}
\end{equation}
The first two terms $\Hamil_{L(R)}=\sum_{k\sigma\in L(R)}\leade{k}\cdagger{k}\cc{k}$ describe the electrons in the leads, where an electron is created (destroyed) in the left $(L)$ or right $(R)$ lead at the energy $\leade{k}$ and spin $\sigma=\up,\down$, by $\cdagger{k}\ (\cc{k})$. The Hamiltonian for a free spin $\bfS$ in the presence of a magnetic field $\bfB_0$ is given by
\begin{equation}
\Hamil_S=-g\mu_B\bfB_0\cdot\bfS,
\label{eq-HS}
\end{equation}
where $g$ and $\mu_B$ are the gyromagnetic ratio and Bohr magneton, respectively. The two leads are weakly coupled via the tunneling Hamiltonian
\begin{equation}
\Hamil_T=\sum_{pq\alpha\beta}\left(\csdagger{p\alpha}
	[T_0\delta_{\alpha\beta}
		+T_1\bfS\cdot\sigma_{\alpha\beta}]\cs{q\beta}+H.c.\right).
\label{eq-HT}
\end{equation}
Here and henceforth, let electrons in the left (right) lead carry the subscript $p\ (q)$. Here also, $\sigma_{\alpha\beta}$ is the vector of Pauli spin matrices, with spin indices $\alpha,\beta$. The rate of the direct tunneling between the leads is denoted by $T_0$ whereas the tunneling with rate $T_1$ is influenced by the presence local spin, both of which are assumed to be spin-independent. Spin-flip transitions in the direct tunneling between the leads are neglected in this model, and it is furthermore implied that the level of the localized electron is far below the Fermi surface of the conduction electrons in both electrodes. Therefore, the voltage bias should not exceed this energy level difference. For convenience the respective amplitudes are taken to be momentum independent (although it is not required). Typically, from the expansion of the work function for tunneling, $T_1/T_0 \sim J/U$, where $J$ and $U$ is the spin-spin exchange interaction parameter and spin-independent tunneling barrier, respectively\cite{balatsky2002,zhu2003} Further, an external magnetic field $\bfB_0(t)$ may applied to the system, see Fig. \ref{fig-system}.

It should be noticed that the model should in principle also contain exchange interaction between the local spin and the electrons in the leads, e.g. terms like $\csdagger{k\alpha}\bfS\cdot\sigma_{\alpha\beta}\cs{k\beta}$, where $k\in L$ or $R$. Such terms will, however, not contribute to the current driven spin dynamics considered here, and are therefore omitted.

Estimates of the signal-to-noise ratio has been presented by Balatsky \etal\cite{balatsky2002} and Nussinov \etal\cite{nussinov2003} for the present model. It was then found that the spectral power density $\overline{\av{I_\omega^2}}\sim I_0^2(T_1/T_0)^2\chi(\omega)$, where $\chi(\omega)$ is a Lorentzian associated with power spectrum of the local spin. The power spectrum of the shot noise is approximately $\av{I^2_{shot}(\omega)}\sim I_0$. Using $I_0\sim1/\tau_e$, where $1/\tau_e$ is the spin-independent scattering rate. Thus, the signal-to-noise ratio becomes $\overline{\av{I^2_\omega}}/\av{I^2_{shot}(\omega)}\sim(T_1/T_0)^2\chi(\omega)/\tau_e$, which in the present system is bounded both from above and below by a number of order unity.\cite{nussinov2003} Hence, the signal may be quite sizable, which enables the further discussions in this paper.

When a time-dependent voltage bias is applied across the tunneling barrier, such that $V(t)=V_{dc}+V_{ac}(t)$, where $V_{dc}$ and $V_{ac}$ are the dc and ac components, a dipole forms around the barrier region through the accumulation or depletion of electron charge. This process results in the time dependence of single-particle energies as $E_{p(q)}=\epsilon_{p(q)}+W_{L(R)}(t)$, with the constraint $W_{L}(t)-W_{R}(t)=eV_{ac}(t)$. However, the occupation of each state in the respective contact remains unchanged and is determined by the distribution established before the time dependence is turned on. Therefore, the chemical potentials on the left $\mu_{L}$ and on the right lead $\mu_{R}$ differs by the dc component of the applied voltage bias, $\mu_{L}-\mu_{R}=eV_{dc}$. The tunnel junction is then characterized by two time scales, the Larmor precession frequency of the spin $\omega_L = g \mu_B|\bfB|$ and the characteristic time scale of the ac field.

The dynamics of the local spin was derived in Ref. \onlinecite{fransson2007} and was found to satisfy the Landau-Lifshitz-Gilbert equation \cite{landau1935,gilbert1955}
\begin{equation}
\frac{d{\bf n}}{dt}=\alpha(t)\frac{d{\bf n}}{dt}\times{\bf n}
	+g\mu_B{\bf n}\times\bfB(t),
\label{eq-Seq}
\end{equation}
under the assumption that the dynamics of the local spin is much slower than the electronic processes. This assumption is motivated by the fact that the energy associated with the spin dynamics, $\hbar\omega_L\sim1\ \mu$eV, is much smaller that the typical electronics energy on the order of 1 meV.\cite{zhu2002} In Eq. (\ref{eq-Seq}) the spin $\bfS(t)=S{\bf n}(t)$, where $S=|\bfS|$, whereas ${\bf n}=\bfS/S$. The damping factor $\alpha(t)\propto ST_1^2/D$, \cite{alpha} at zero temperature, where $2D$ is the band widths of the leads,\cite{fransson2007} and it is reasonable to believe this is the case also for finite temperatures. Thus, by assuming large $D$, the damping $\alpha$ becomes negligible. 

The effective magnetic field $\bfB(t)=\bfB_0(t)+\bfB_\text{ind}^{(1)}(t)+\bfB_\text{ind}^{(2)}(t)$, where the induced magnetic fields can be expressed as\cite{fransson2007}
\begin{eqnarray}
\lefteqn{
g\mu_B\bfB_\text{ind}^{(1)}(t)=
	-2T_0T_1\sum_{pq\sigma}\sigma_{\sigma\sigma}^z
		[f(\xi_{p\sigma})-f(\xi_{q\sigma})]
}
\nonumber\\&&\times
		\im\int_{-\infty}^t
			e^{i[(\xi_{p\sigma}-\xi_{q\sigma})(t-t')+\varphi(t)-\varphi(t')]}
		dt'
{\hat{\bf z}}
\label{eq-B1}
\end{eqnarray}
\begin{eqnarray}
g\mu_B\bfB_\text{ind}^{(2)}(t)&=&
	S\int_{-\infty}^t \Bigl\{
	K_{xy}^{(2)}(t,t')(n_y\hat{\bf x}-n_x\hat{\bf y})
\nonumber\\&&
	+[K_{xx}^{(2)}(t,t')-K_{zz}^{(2)}(t,t')]n_z\hat{\bf z}\Bigr\}dt'
\label{eq-B2}
\end{eqnarray}
where
\begin{eqnarray}
K_{xy}^{(2)}(t,t')&=&2\pi T_1^2
	\re\sum_{pq\sigma}[f(\xi_{p\sigma})-f(\xi_{q\bar\sigma})]
\nonumber\\&&\times
			e^{i[(\xi_{p\sigma}-\xi_{q\bar\sigma})(t-t')+\varphi(t)-\varphi(t')]}
\label{eq-Kxy}
\end{eqnarray}
and 
\begin{eqnarray}
\lefteqn{
K_{xx}^{(2)}(t,t')-K_{zz}^{(2)}(t,t')=}
\nonumber\\&&
	-2\pi T_1^2\im\sum_{pq\sigma}
			\Bigl\{[f(\xi_{p\sigma})-f(\xi_{q\bar\sigma})]
			e^{i(\xi_{p\sigma}-\xi_{q\bar\sigma})(t-t')}
\nonumber\\&&
	-[f(\xi_{p\bar\sigma})-f(\xi_{q\sigma})]
			e^{i(\xi_{p\bar\sigma}-\xi_{q\sigma})(t-t')}\Bigr\}e^{i[\varphi(t)-\varphi(t')]}
\label{eq-Kxz}
\end{eqnarray}
Here $\varphi(t)=\int_{t_0}^tV(t')dt'$, for some initial time $t_0$, such that $\varphi(t)-\varphi(t')=\int_{t'}^tV(t'')dt''$, whereas $\xi_{p(q)\sigma}=\leade{p(q)}+\mu_{L(R)}$.

These induced fields are direct responses to the biasing of the spin-polarized leads. The spin-imbalances ($N_{L(R)\up}-N_{L(R)\down}$) in the leads generate a field along the $z$-direction which modifies the Larmor frequency of the local spin. More important for the switching, however, is whether or not the magnetic moments of the leads are equal and/or parallel. Unequal and/or non-parallel magnetic moments of the leads generate a magnetic field which tends to force the local spin into the magnetic direction of the source lead, as is shown below.

\begin{figure}[t]
\begin{center}
\includegraphics[width=8.5cm]{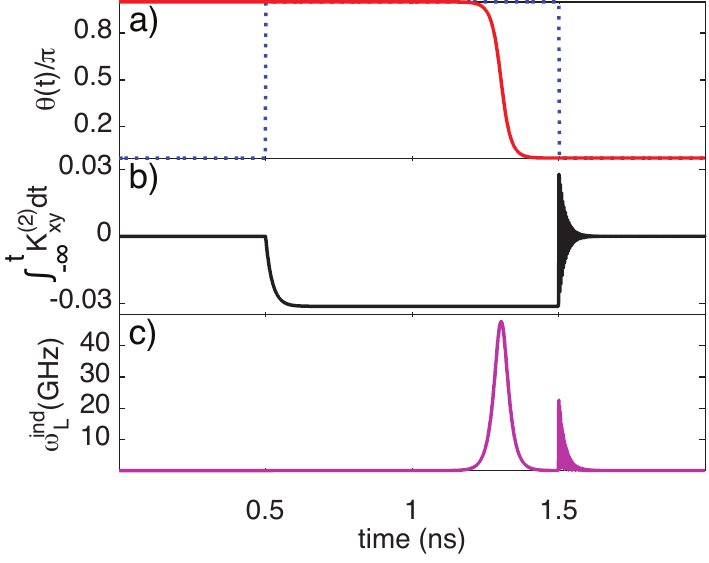}
\end{center}
\caption{(Color online) Time-dependent variation of a) $\theta(t)$, b) $\int_{-\infty}^tK_{xy}^{(2)}dt'$, and c) the induced Larmor frequency $\omega_L^\text{ind}(t)$, for a square bias voltage pulse, c.f. dotted line in a). Here, $V_{ac}=1$ mV, $T_1N_0=0.1$, $p_L=-p_R=1/2$.}
\label{fig-onset}
\end{figure}

In order to better understand the motion of the local spin, consider the corresponding classical equation of motion, which is consistent with the parametrization of the spin on the unit sphere, i.e. $S{\bf n}=S(\cos{\phi}\sin{\theta},\sin{\phi}\sin{\theta},\cos{\theta})$. With $\bfB_0=B_0\hat{\bf z}$ time-independent, the equations can be written
\begin{equation}
\begin{array}{rcl}
\displaystyle \frac{d\phi}{dt} & = &
	\displaystyle - g\mu_B[B_0+B_\text{ind}^{(1)}(t)]
\\& &	\displaystyle
	+S\int_{-\infty}^t[K_{xx}^{(2)}(t,t')-K_{zz}^{(2)}(t,t')]dt'\cos{\theta},
\\\\
\displaystyle \frac{d\theta}{dt} & = &
	\displaystyle S\int_{-\infty}^t K_{xy}^{(2)}(t,t')dt'\sin{\theta}.
\end{array}
\label{eq-angles}
\end{equation}
The equation for the polar angle $\theta$ can be analytically solved and in the stationary regime it gives
\begin{equation}
\theta(t)=2\arctan{\biggl(\tan{\frac{\theta_0}{2}}
	e^{-2\pi eV_{dc}T_1^2\sum_\sigma\sigma_{\sigma\sigma}^z
		N_{L\sigma}N_{R\bar\sigma}(t-t_0)}\biggr)},
\label{eq-stattheta}
\end{equation}
where $\theta_0=\theta(t_0)$. This expression clearly shows the possibility to switch the local spin by means of a stationary bias voltage. For example, configuring the leads such that $N_{L\up}N_{R\down}-N_{L\down}N_{R\up}>0$,\cite{magn_interpret} and biasing the system by $V_{dc}>0$, gives an exponential decay of the argument which leads to that the polar angle $\theta(t)\rightarrow0$ (local spin becomes $\up$), as $t\rightarrow\infty$, for any given initial polar angle $\theta_0$. By the same argument $\theta(t)\rightarrow\pi$ (local spin becomes $\down$) if the bias $V_{dc}<0$ is applied.

Using the result for the polar angle within the stationary regime, the characteristic time-scale $\hbar/\tau_c\simeq2\pi eVT_1^2\sum_\sigma\sigma_{\sigma\sigma}^zN_{L\sigma}N_{R\bar\sigma}$ for the polar angle motion of the spin is obtained. Parametrizing the spin-polarized DOS $N_{L(R)\sigma}=N_0(1+\sigma p_{L(R)})/2$,\cite{fransson2005} where $-1\leq p_{L(R)}\leq1$, assuming $T_1N_0\sim0.1$, $p_L=-p_R=-1/2$, and $V\sim1$ mV gives the characteristic time scale $\tau_c\approx5$ ps, which is sufficiently short to switch the spin from being $\down$ to $\up$ within 1 ns, see Fig. \ref{fig-onset} a).

In a realistic set-up of the system it would be desirable to obtain the current induced switching by application of bias pulses of some time span $\tau_s$. A sudden onset of a bias pulse leads to transient effects in the induced magnetic fields $\bfB_\text{ind}^{(n)}(t),\ n=1,2$, which transfer into the motion of the spin. Therefore, consider application of the bias $V_{dc}+V_{ac}[\Theta(t-\tau_0)-\Theta(t-\tau_1)]$, where $\tau_1=\tau_0+\tau_s$. We also assume that $\bfB_0=0$ since we are interested in the local spin dynamics generated by the spin-polarized current. As shown above, a stationary bias will eventually line up the local spin in the magnetic direction of the source lead and since the main interest lies in a switching obtained by a pulsed bias, henceforth $V_{dc}=0$. Then, the equation of motion for $\theta(t)$ can be written
\begin{eqnarray}
\frac{d\theta}{dt}&=&
	-2\pi ST_1^2
	\sum_\sigma\sigma_{\sigma\sigma}^zN_{L\sigma}N_{R\bar\sigma}
	eV_{ac}\biggl\{\biggl[1-e^{-(t-\tau_0)/\tau}\biggr]
\nonumber\\&&\times
		[\Theta(t-\tau_0)-\Theta(t-\tau_1)]
		+\biggl[e^{-(t-\tau_1)/\tau}
\label{eq-St}\\&&
			-e^{-(t-\tau_0)/\tau}\biggr]\cos{[eV_{ac}(t-\tau_1)]}\Theta(t-\tau_1)
		\biggr\}\sin\theta
\nonumber
\end{eqnarray}
where the time-scale $\tau$ relates to the electronic tunneling processes. Such processes are of the order of 1 fs, which is much faster than the characteristic time-scale $\tau_c$ and the Larmor frequency $\hbar\omega_L$ for biases between 1 - 100 mV.\cite{zhu2002} For e.g. a 1 ns bias pulse of 1 mV, the critical time scale is $\sim100$ fs, which may be achieved within present experimental state-of-the-art-technology for nanoscale systems. Physically the time-scale $\tau$ implies that the induced magnetic field is a retarded response to the bias across the junction.

The calculated time-dependence of the polar angle is plotted in Fig. \ref{fig-onset} a) for a square bias voltage pulse, and it is readily seen that the angle goes from $\pi$ to 0 within the time span of the pulse, hence, the pulse is sufficiently long to flip the spin from $\down$ to $\up$. The plot in Fig. \ref{fig-onset} b) shows the induced magnetic field which is acting to align the local spin along the magnetic direction of the source lead. At the onset (termination) of the bias pulse, the amplitude of the induced magnetic field grows (decays) exponentially, as expected from Eq. (\ref{eq-St}). At the pulse termination, however, there are additional oscillations in the induced field, as a reaction to the removed bias. These oscillations, are not visible in the motion of the polar angle, since they are exponentially suppressed, c.f. Eq. (\ref{eq-stattheta}).

The Larmor frequency $\omega_L$ of the precession of the local spin is affected by the time-dependent variation of $\theta$ that gives rise to a momentary change of its value. By definition
\begin{eqnarray}
\lefteqn{
\omega_L^2(t)\equiv(g\mu_B)^2|\bfB(t)|^2=
	\biggl(B_0+B_\text{ind}^{(1)}(t)
}
\nonumber\\&&
		+S\int_{-\infty}^t[K_{xx}^{(2)}(t,t')-K_{zz}^{(2)}(t,t')]dt'\cos\theta(t)
	\biggr)^2.
\nonumber\\&&
	+\biggl(S\int_{-\infty}^tK_{xy}^{(2)}(t,t')dt'\biggr)^2\sin^2\theta(t)
\label{eq-larmor}
\end{eqnarray}
The time-dependent parts of the induced fields in the first term are equal and can be written as $(V_{dc}=0$)
\begin{equation}
eV_{ac}\biggl[e^{-(t-\tau_0)/\tau}-e^{-(t-\tau_1)/\tau}\biggr]
	\sin{[eV_{ac}(t-\tau_1)]}\Theta(t-\tau_1).
\end{equation}
Hence, this field does not affect the Larmor frequency until after the bias pulse has terminated, and then it generates an oscillatory variation of $\omega_L$ with an exponential decay, see Fig. \ref{fig-onset} c). These oscillations are sufficiently large to enable read-out of the time instant when the pulse is terminated.

\begin{figure}[t]
\begin{center}
\includegraphics[width=8.5cm]{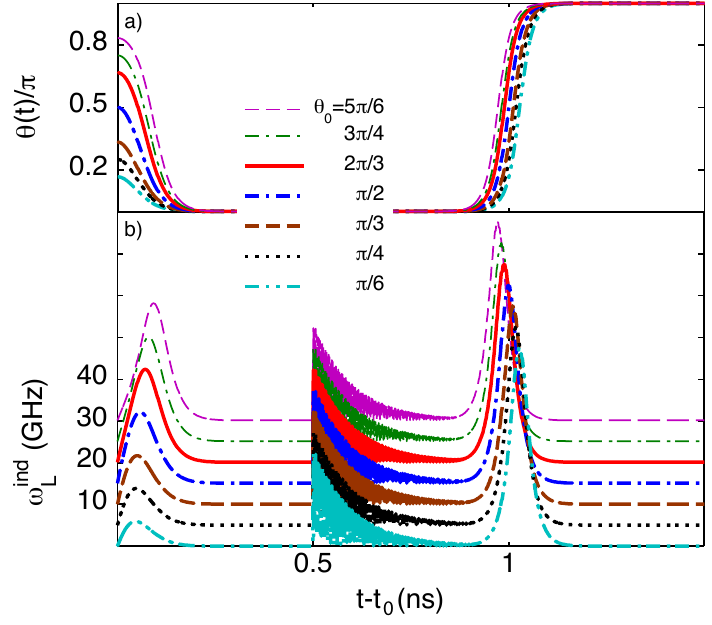}
\end{center}
\caption{(Color online) Time-dependent variation of a) $\theta(t)$  and b) $\omega_L^\text{ind}(t)$, for different initial polar angles $\theta_0$, under the bias $V(t)=V_{dc}+V_{ac}\{[\Theta(t-t_0)-\Theta(t-t_1)]-[\Theta(t-t_1)-\Theta(t-t_2)]\}$, with $t_1-t_0=t_2-t_1=1$ ns. For clarity, the plots in panel b) have been shifted 5 GHz upwards for increasing $\theta_0$. Other parameters are as in Fig. \ref{fig-onset}.}
\label{fig-onset_theta0}
\end{figure}

The last term in Eq. (\ref{eq-larmor}), containing the field that generates the spin-flip, vanishes as long as $\theta(t)=0,\pm\pi$, that is, when the spin is either $\up$ or $\down$ in the global spin quantization axis. This term gives a contribution, however, during the short time interval when the spin flips, since then $0<|\theta(t)|<\pi$. Hence, one would expect a sharp peak in the time-evolution of $\omega_L(t)$, which indeed is seen in Fig. \ref{fig-onset} c). This peak in the time-evolution of the Larmor frequency is a fingerprint of an actual occurrence of a spin-flip and would be measurable. Using the values comparable to those in Fig. \ref{fig-onset}, results in an amplitude of the peak of the order of 50-500 GHz in case of vanishing external magnetic field, for biases 1 - 10 mV.

In the discussion we have neglected influences from e.g. anisotropy fields. The absence of such or similar fields that would stabilize the spin direction, cause the spin orientation to drift randomly in equilibrium. It has been shown that the spin-polarized current between two ferromagnetic leads has such effect on the local spin, i.e. the effect of the spin-polarized current on the local spin is to stabilize its orientation. The plots in Fig. \ref{fig-onset_theta0} a) show $\theta(t)$ for several different initial polar angles $\theta_0$, given the bias $V(t)=V_{dc}+V_{ac}\{[\Theta(t-t_0)-\Theta(t-t_1)]-[\Theta(t-t_1)-\Theta(t-t_2)]\}$, with $t_1-t_0=t_2-t_1=1$ ns. The plots illustrate how the spin-polarized current forces the local spin to be aligned with the majority spin of the source lead, $\theta(t)\rightarrow0$ for $t_0\leq t\leq t_1$. Furthermore, by reversing the bias voltage, the spin orientation can be reversed at will, which is illustrated by the plots in Fig. \ref{fig-onset_theta0} a) where $\theta(t)\rightarrow\pi$, in the interval $t_1\leq t\leq t_2$. Panel b) in Fig. \ref{fig-onset_theta0} further illustrates how the Larmor frequency peaks at the instant of the spin-flip, both when its polar angle approaches zero and $\pi$. The increasing amplitude of the peak with increasing difference between the final and initial polar angle is expected, because of the increasing reorientation path the spin has to traverse.

The result displayed in Eq. (\ref{eq-St}) shows that it would be possible to flip the spin by having only one of the leads ferromagnetic and the other non-magnetic. Thus, a set-up of an scanning tunneling microscope (STM) with a spin-polarized STM tip and non-magnetic substrate surface would correspond to having, say, $p_L\neq0$ and $p_R=0$. Hence, letting $p_L=1/2$ and using the same parameters as in Fig. \ref{fig-onset}, one finds a time scale for the spin flip of roughly $\tau_c\simeq10$ ps, that is, doubled the time scale compared to the case discussed above. The spin flip effect should nonetheless be likewise observable as in the case of two ferromagnetic leads in anti-parallel alignment, however, on a larger time scale.

The result in Eq. (\ref{eq-St}) also shows that if both leads are being ferromagnetic and in parallel alignment, the resulting spin flip effect on the local spin decreases since the induced fields from the left and right leads tend to cancel each other. The polar angle becomes a constant of motion whenever the two leads are equally strong ferromagnets in parallel configuration, since then $p_L=p_R=p$ which yields $\sum_\sigma \sigma_{\sigma\sigma}^zN_{L\sigma}N_{R\bar\sigma}=(N_0/2)^2[(1+p)(1-p)-(1-p)(1+p)]=0$.

Anisotropy field acting on the local spin has been omitted in the present paper for simplicity. In a realistic system, however, the local spin is likely subject to some type of anisotropy field which stabilize definite spin directions. Such fields may be small enough to not substantially increase the time scale of the spin-flip processes in the present situation. Without any type of anisotropic field acting on the local spin, its orientation will in equilibrium, i.e. $V(t)=0$, most likely describe a random drift due to exchange interaction with equilibrium fluctuations in the current between the leads. Such a current will lack any definite spin-polarization that the non-equilibrium situation can present.

In summary, it is demonstrated that the dynamics of a local spin embedded in the tunnel junction between ferromagnetic leads can be manipulated at will by means of a time-dependent bias voltage. Especially, spin-flip transitions can be stimulated at a sub-nanosecond time-scale by bias voltage pulses, even for a moderate spin-polarization of the leads. The time-evolution of the Larmor frequency of the local spin sharply peaks at the spin-flip event, something that determines actual occurrence of a spin-flip. These findings would be useful in the advent spintronics and quantum information technology.

The author thanks A. V. Balatsky and J. -X. Zhu for useful discussions. This work has been supported by US DOE, LDRD and BES, and was carried out under the auspices of the NNSA of the US DOE at LANL under Contract No. DE-AC52-06NA25396.


\begin{thebibliography}{99}
\bibitem{magn_interpret} $\sum_\sigma\sigma_{\sigma\sigma}^zN_{L\sigma}N_{R\bar\sigma}>0$ either means i) both leads being ferromagnetic in a non-parallel configuration , or ii) only left (right) lead is ferromagnetic. In both cases there is a spin $\up$ ($\down$) majority in the left (right) lead.
\end{thebibliography}
\end{document}